\documentclass[aps,prd,nofootinbib,notitlepage]{revtex4-1}

\input glyphtounicode
\usepackage[T1]{fontenc}
\usepackage[utf8]{inputenc}

\usepackage[english]{babel}

\usepackage{amssymb,amsfonts,mathtools}
\usepackage{amsthm}
\usepackage{enumerate}
\usepackage{soul}
\setstcolor{red}

\usepackage[dvipsnames]{xcolor}
\definecolor{hypercolor}{rgb}{0,0.2,0.7}
\usepackage[breaklinks,final]{hyperref}

\usepackage{footmisc}

\usepackage{graphicx}
\usepackage{subcaption}

\newcommand{\defn}{\coloneqq}

\usepackage{upgreek}
\newcommand{\dif}{\mathrm{d}}

\newcommand{\conlaw}{\mathop{\#}\nolimits}

\theoremstyle{definition}

\hypersetup{
	unicode,
	colorlinks,
	linkcolor=hypercolor,
	urlcolor=hypercolor,
	citecolor=hypercolor,
	bookmarksnumbered,
	bookmarksdepth=2,
	pdfauthor={Christian Pfeifer},
	pdftitle={...}
}

\begin{document}

\title
{
Observers' measurements in premetric electrodynamics I: Time and radar length
}

\author{Norman G\"urlebeck}
\email{norman.guerlebeck@zarm.uni-bremen.de}
\affiliation{ZARM, University of Bremen, Am Fallturm 2, 28359 Bremen, Germany.
}

\author{Christian Pfeifer}
\email{christian.pfeifer@ut.ee}
\affiliation{Laboratory of Theoretical Physics, Institute of Physics, University of Tartu, W. Ostwaldi 1, 50411 Tartu, Estonia.}

\begin{abstract}
The description of an observer's measurement in general relativity and the standard model of particle physics is closely related to the spacetime metric. In order to understand and interpret measurements, which test the metric structure of the spacetime, like the classical Michelson–Morley, Ives-Stilwell, Kennedy-Thorndike experiments or frequency comparison experiments in general, it is necessary to describe them in theories, which go beyond the Lorentzian metric structure. However, this requires a description of an observer's measurement without relying on a metric. We provide such a description of an observer's measurement of the fundamental quantities time and length derived from a premetric perturbation of Maxwell's electrodynamics and a discussion on how these measurements influence classical relativistic observables like time dilation and length contraction. Most importantly, we find that the modification of electrodynamics influences the measurements at two instances: the propagation of light is altered as well as the observer's proper time normalization. When interpreting the results of a specific experiment, both effects cannot be disentangled, in general, and have to be taken into account. 
\end{abstract}

\maketitle


\section{Introduction}
In special and general relativity, physical measurements of an observer are closely related to the Lorentzian spacetime metric~$g$. An observer propagating through spacetime on a timelike worldline $\gamma$ possesses a clock, which shows its proper time and a set of three rulers, which show the observer's orthogonal spatial unit length directions. An observer employs these unit directions to interpret physical observables with respect to its rest-frame. Mathematically, they are modeled by an orthonormal frame $\{e_\mu\}_{\mu=0}^{3}$ of the metric~$g$, i.e. $g(e_\mu,e_\nu) = \eta_{\mu\nu}$. The observer's clock, or unit time direction, is given by $e_0 = \dot \gamma$ while the spatial directions $e_\alpha, \alpha=1,2,3$ are interpreted as the unit space rulers of the observer. The just mentioned physical interpretation of the frames of the metric can be derived operationally in terms of radar signal experiments, as is for example nicely demonstrated in the textbook \cite[Sec.\ 9]{DfC}.

A rarely discussed question is what happens to this description of observers and their measurements when the theory, whose observables one investigates is not based on a Lorentzian spacetime metric defining the geometry of the spacetime manifold. 
In physics, such theories appear for example in the axiomatic approach to electrodynamics as pre-metric or area metric electrodynamics \cite{Hehl,Schuller:2009hn}, in the Standard Model Extension \footnote{We will highlight in Sec.\ \ref{sec:GLE} that the photon sector of the SME is included in the premetric approach to electrodynamics.}(SME) \cite{Colladay:1998fq,Kostelecky:2002hh,Bluhm:2005uj}, the Robertson-Mansouri-Sexl\footnote{The "metric" of that theory does not solely depend on the manifold's coordinates but also on the frame it is measured in, in particular, its motion with respect to a preferred frame.}(RMS) theory \cite{Robertson1949,Mansouri1977,Mansouri1977a,Mansouri1977b,Lammerzahl2002}, quantum gravity phenomenology inspired $\kappa$-Poincare invariant field theories \cite{Barcaroli:2017gvg} or in general field theories and models which lead to a dispersion relation that differs from the general relativistic one of the standard model of particle physics on curved spacetime. The question is how are the unit length and unit time directions of an observer identified operationally for such theories, what are an observer's equal time surfaces and how do observers measure spatial lengths? In special and general relativity all these different concepts are directly determined by the spacetime metric. In this article we will review, collect and construct the mathematical objects to describe the aforementioned physical concepts consistently starting from from a general local and linear modification of Maxwell vacuum electrodynamics on Minkowski spacetime.

A partial answer to the description of observers without a metric was given in \cite{Raetzel:2010je}, where a generalized notion of timelike observer worldlines and the observer's time measurements are identified from general dispersion relations. The latter are obtained as level sets of Hamilton functions on the cotangent bundle of the manifold where physics takes place, which describe the geometric optic limit of the field theories in consideration. Passing from the Hamiltonian to the Lagrangian description, it became clear that in general the motion of massless (light) and massive (observer) particles have to be described by two different Lagrange functions. For standard model field theories on curved spacetime and general relativity, these length functions are closely related and are given by the metric length of tangent vectors, respectively. However, in general observers and their measurements need to be described in terms of the two different Lagrange functions. An example where the influence of the non-equal light and massive particle/observer Lagrangian has been taken into account explicitly is the derivation of quantum-energy-inequalities in premetric electrodynamics inside an uniaxial crystal \cite{Fewster:2017mtt}.

What still has to be done in the analysis of observers on the basis of non-metric field theories is the derivation of further observable consequences such as the most common relativistic effects: an observer's measurement of time dilation and length contraction. These and other relativistic observables rely on an observer's measurement of spatial lengths and time intervals. The main goal of this article is to give an \emph{operational} description of the physical spatial length an observer measures in terms of a radar experiment. This extends the previous analysis of the notion of spatial length on Finsler spacetimes \cite{Pfeifer:2014yua} to general dispersion relations. We like to stress already, that the radar length we obtain depends on both of the two different aforementioned Lagrangians.

The influence of the observer model and of an observer's measurement of radar lengths on the prediction of observables is particularly relevant for the classical tests of special relativity and local Lorentz invariance \cite{Tobar2010,Nagel2015,Botermann2014}: the Michelson–Morley experiment, the Ives-Stilwell experiment and the Kennedy-Thorndike experiment. To constraint deviations from special relativity and local Lorentz invariance these experiments are often modelled in the SME or RMS framework. The first approach considers very general parametrized modifications of the Lagrangian of the standard model and looks for phenomenological consequences to constrain the parameters of these modifications. The second assumes that observers are related by modified Lorentz transformations and constraints for these modifications are deduced from experiments. In view of proposed satellite based versions of these experiments \cite{Dittus2007,Gurlebeck2015,Gurlebeck2017}, the interpretation of the measurements requires a clear theoretical background.

Here a missing link is a consistent derivation of the notion of observers and their measurements from the underlying field theories that describe the physics involved in the experiment. Such a procedure connects the field theoretical (SME-like) and the kinematical (RMS-like) approaches to observables from non local Lorentz invariant or non-metric field theories. We close this gap by combining an observer's measurement of time, as outlined in \cite{Raetzel:2010je}, with our new derivation of an observer's measurement of length both based on the geometric optic limit of the field theories considered.

The presentation of our result is structured as follows: 
We will start in Sec.\ \ref{sec:GLE} and Sec.\ \ref{sec:PartProp} by considering a first order perturbation of Maxwell vacuum electrodynamics in a Minkowski spacetime in the framework of premetric electrodynamics, which contains the minimal extension of the electrodynamics sector used in the SME, where the Lagrangian is quadratic in the field strength tensor and does not explicitly depend on the $4$-form potential. In the latter Section, we apply the general methods of \cite{Raetzel:2010je} to the theory in order to derive the Lagrangians, which govern the motion of massless (light) and massive (observer) particles. It turns out these are two different functions, the latter being a Finsler function. This reveals a yet unobserved relation between the minimal SME electrodynamics and Finsler geometric structures and extends earlier connections between Finsler geometry and the SME, see e.g.\ \cite{Kostelecky:2011qz}.

Next, we use our knowledge about the motion of light and observers in Sec.\ \ref{sec:timelength} to define an observer's measurement of time by employing the clock postulate. It states that the time an observer measures between two events is given by the length of the observer's worldline connecting these events. The length must be measured with the Lagrange function that governs the motion of massive particles and determines an observer's equal time surfaces which give rise to the observer's spatial directions. The central result of our analysis is the derivation of the radar length from a radar experiment. This radar length assigns physical operationally accessible numbers to distances relative to an observer and is the basic quantity that describes interferometric experiments. Eventually, we briefly discuss the classical relativistic observables time dilation and length contraction in Sec.\ \ref{sec:classrel} on the basis of the time measuring Lagrangian and the radar length we derived before.

The explicit calculations are done for perturbative theory on flat spacetime, respectively infinitesimal on curved spacetime, in order to display explicit formulas and to give a clear impression how our derivation works. This is also physically justified, where we expect any deviation from our current metric theories to be small in recent experiments. Throughout this perturbative approach another important subtlety worth mentioning appears. To derive first order corrections to the notion of radar length in special and general relativity from a first order perturbation of Maxwell's electrodynamics, it is necessary to derive the geometric optic limit of the theory to second order in the perturbation. Otherwise the desired corrections cannot be determined.

The same analysis on curved spacetime will yield corrections to the derived formulas in terms of the relevant spacetime curvature similar to the curvature corrections to the special relativistic expression in general relativity \cite[Sec.\ 9]{DfC} and is a project which builds on the results obtained here.

An interesting aspect of our approach to the description of an observer's measurement appears when one reviews the two fundamental postulates of special relativity in the light of our results. We leave the first postulate, the clock postulate, untouched, while the second postulate, the constancy of the speed of light, becomes derivable from a theory of electrodynamics, which predicts the motion of light. The local spatial speed of light an observer measures, can then be derived, and, if it is constant or not, depends on the theory of electrodynamics employed. Thus, instead of using the second postulate of special relativity, we rather formulate a theory of electrodynamics and derive the relativity theory from it. In this spirit, Maxwell's electrodynamics implies the constancy of the speed of light and special relativity much like in the historic derivation of special relativity as can be nicely recognized from the original papers collected in \cite{ThePrincipalOfRelativity}. Similarly, deviations of the former lead to modifications of the latter. 

The notational and sign conventions in this article are: The Minkowski metric $\eta$ has signature $(+,-,-,-)$, symmetrization $T{}^{(a_1....a_n)}$ and antisymmetrization $T{}^{[a_1....a_n]}$ brackets over $n$ indices are defined with a factor $(n!)^{-1}$.

\section{Premetric electrodynamics}\label{sec:GLE}
To derive the desired relativistic observables from electrodynamics on a non-metric background in a covariant way, we will use the language of general linear or \emph{premetric} electrodynamics, which we briefly recall here. Electrodynamics is described by Maxwell's equations in terms of the electromagnetic field strength $2$-form $F$, the excitation $2$-form $H$ and a conserved source current $3$-form~$J$
\begin{align}\label{eq:maxwell}
	\dif F=0,\ \dif H = J.
\end{align}
On its own, these equations are not predictive but require a so-called constitutive relation $\conlaw$. It defines $H$ as a function of $F$. All theories defined by the above equations and a constitutive relation can be considered as a theory of electrodynamics \cite{Hehl}.

The theories of electrodynamics constructed with a local and linear, also called premetric, constitutive relation are of particular interest:
\begin{align}\label{eg:genlinedyn}
	H = \conlaw(F),\quad H_{ab} = \frac{1}{2}\kappa_{ab}{}^{cd}F_{cd} = \frac{1}{4} \epsilon_{abmn}\chi^{mncd}F_{cd}.
\end{align}
$\kappa$ is the local constitutive tensor and $\chi$ the so-called constitutive density. In local coordinates, the dynamical equations~\eqref{eq:maxwell} of the theory become
\begin{equation}\label{eq:genlincoord}
\frac{1}{2}\partial_{[a} \big( \kappa_{bc]}{}^{de} \partial_d A_e \big)
= \frac{1}{4} \partial_{[a} \big( \varepsilon_{bc]de} \chi^{defg} \partial_f A_g \big)
= \frac{1}{3!} J_{abc}\,,
\end{equation}
or in terms of the field strength
\begin{align}\label{eq:maxwellgled}
	F_{ab} = 2 ( \partial_a A_b - \partial_b A_a ),\quad \partial_b( \chi^{abcd}F_{cd}) = j^a = \frac{1}{3!}\epsilon^{abcd}J_{abcd}\,.
\end{align}
Eq.\ \eqref{eq:maxwellgled} is obtained from Eq.\ \eqref{eq:genlincoord} by contraction with the total antisymmetric Levi-Civita symbol $\epsilon^{abcd}$, which is normalized such that $\epsilon^{0123}=1$.
The covariant formulation of premetric electrodynamics enables a precise study of the propagation behaviour of light in this general setting \cite{Rubilar:2007qm} and for the quantization of the theory \cite{Rivera:2011rx, Pfeifer:2016har,Fewster:2017mtt}.

General premetric electrodynamics covers a huge variety of effects. Interesting examples are vacuum electrodynamics and quantum corrections to electrodynamics from renormalization on curved spacetimes \cite{Drummond}, electrodynamics in media~\cite{Hehl} and area-metric electrodynamics~\cite{Punzi:2007di,Punzi:2007di,Schuller:2009hn}. In the first case, the constitutive law is defined solely through the spacetime metric, in the second by the metric and its curvature, in the third by the properties of the medium in consideration and in the fourth case by a general area measure on the manifold.

The geometric optical limit of the theory is described by the Fresnel polynomial \cite{Rubilar:2007qm}
\begin{align}\label{eq:Fresnel}
\begin{split}
\mathcal{G}(k) & = G^{abcd}k_ak_bk_ck_d,\\
G^{abcd} & = \frac{1}{4!}\epsilon_{e_1e_2e_3e_4}\epsilon_{e_5e_6e_7e_8}\chi^{e_2e_1e_6 (a}\chi^{b|e_3 e_7|c}\chi^{d)e_4 e_8e_5}.
\end{split}
\end{align}
It determines the motion of particles via Hamilton's equations of motion and its level sets represent the dispersion relation of the theory, i.e., $\mathcal{G}(k)=m^4$ for massive and $\mathcal{G}(k)=0$ for massless particles. The Fresnel polynomial determines the causal structure of the theory, which can be much more involved than a metric light cone. It may contain for example several light cones as it is the case for birefringent crystals \cite{Perlick}.

\section{Particle propagation from weak premetric electrodynamics}\label{sec:PartProp}
Our aim is to derive an observer's measurement of length from a radar experiment in the next section. The experiment consists of an observer, who sends out light rays which get reflected and propagate back to the observer. As a prerequisite, we derive here the motion of observers and light rays predicted by the most general position independent local and linear first order perturbation of Minkowski spacetime Maxwell- or \emph{metric electrodynamics}, which can be derived from an action. We call this theory, in line with the definition of premetric electrodynamics in the previous section, \emph{weak premetric electrodynamics} or short WPE. It includes the theory of electrodynamics employed in the minimal SME \cite{Kostelecky:2002hh,Kostelecky:2009zp}, and the theories of electrodynamics from which the modified spectrum of the hydrogen atom has been derived in \cite{Drummond:2016ukf,Grosse-Holz:2017rdt}.

We derive the Fresnel Polynomial, which determines the propagation of particles and the causal structure of the spacetime manifold and translate the Hamiltonian description of particle propagation based on the Fresnel Polynomial into its corresponding Lagrangian, respectively Finslerian formulation.

\subsection{The geometric optics limit}
The most general WPE is described by a constitutive density of the form
\begin{align}\label{eq:constdensity}
\chi^{abcd} =|\det(\eta)|^{\frac{1}{2}} (2 \eta^{a[d}\eta^{c]b}+\mathcal{K}^{abcd}),
\end{align}
where the term $\mathcal{K}^{abcd}$ parametrizes the deviations from metric electrodynamics and has the following properties $\mathcal{K}^{abcd}=\mathcal{K}^{[ab][cd]}=\mathcal{K}^{[cd][ab]}$ and $\mathcal{K}^{a[bcd]}=0$, i.e., it has the same symmetries as the Riemann curvature tensor. 
The pair-symmetry excludes dissipative energy-loss effects induced by the modified background of the theory, while the Bianchi-Identity symmetry excludes an axion scalar in the background which does not contribute to the electromagnetic energy-momentum nor influence the propagation of light.\footnote{For further details on the decomposition of the constitutive density into a symmetric trace-free, antisymmetric (skewon) and trace (axion) part, see \cite{Hehl}.}
Throughout this article we consider the constitutive density in a Cartesian coordinate system, in which the determinant of the Minkowski metric is one and the $\mathcal{K}^{abcd}$ are assumed to be constant. We assume the existence of such a coordinate system over the region of spacetime, where the radar experiment is performed, i.e., our results hold locally on curved or globally on flat spacetime.

Evaluating the second field equation in \eqref{eq:maxwell} with the constitutive law \eqref{eq:constdensity}, we get
\begin{align}\label{eq:edynsme}
	(\eta^{ac}\eta^{bd} + \mathcal{K}^{abcd})\partial_bF_{cd}=0.
\end{align}
This is exactly the equation of motion employed in the minimal SME electrodynamics \cite{Kostelecky:2002hh,Kostelecky:2009zp} with the difference that there $\mathcal{K}^{abcd}$ is additionally assumed to be double trace free $\mathcal{K}^{abcd}\eta_{ac}\eta_{bd}=0$.

To obtain the desired observables from the field theory to first order in the perturbation parameters~$\mathcal{K}^{abcd}$, it is necessary to derive the Fresnel polynomial to second-order in these. In particular, the lightlike directions we seek to determine to first order depend on the root of the second order geometric optic limit as we demonstrate in~\eqref{eq:lightlike}. In the framework of the SME, the same feature is present, see \cite[Eq. (4)]{Kostelecky:2001mb}. Nonetheless, the second order Fresnel polynomial \eqref{eq:FresnelPert} derived from \eqref{eq:edynsme}, whose level sets define the dispersion relation for timelike and lightlike covectors in that theory, we display here for the first time to our knowledge
\begin{align}\label{eq:FresnelPert}
	\mathcal{G}(k) = \eta^{-1}(k,k)^2 - \eta^{-1}(k,k)\mathcal{K}(k,k) + \frac{1}{2}\big(\mathcal{K}(k,k)^2 -\mathcal{J}(k,k,k,k) \big) + O(\mathcal{K}^3),
\end{align}
where $\mathcal{K}(k,k)$ is defined as
\begin{align}\label{eq:RicciK}
\mathcal{K}(k,k) = \mathcal{K}^{ac}k_ak_c,\quad\text{with}\quad \mathcal{K}^{ac}= \mathcal{K}^{abcd}{}\eta_{bd} = \mathcal{K}^{abc}{}_{b}.
\end{align} 
Indices are raised and lowered with the Minkowski metric $\eta$ although $\eta$ is not a fundamental object in the theory we are investigating but rather one of several building blocks. In the second order perturbation of $\mathcal{G}$, the symbol $\mathcal{J}(k,k,k,k)$ denotes a symmetric fourth rank tensor contracted with four covectors $k$, which is build from a particular square of the coefficients $\mathcal{K}^{abcd}$
\begin{align}\label{eq:KKtensor}
\mathcal{J}(k,k,k,k) = \mathcal{J}^{acef}k_ak_ck_ek_f = \mathcal{K}^{abcd}\mathcal{K}^{e}{}_{b}{}^f{}_d k_a k_c k_ek_f.
\end{align}
As usual, $O(\mathcal{K}^n)$ denotes terms at least of the order $n$ in $\mathcal{K}^{abcd}$.

The observables we seek to derive are best described in configuration space and not on phase space. Thus we need to determine the directions along which light propagates and a precise notion of directions along which observers travel as well as their proper-time parametrization. This can be done by deriving the point-particle action for massless and massive trajectories, which are defined by Lagrange functions we call $L^\#$ and $L^*$, respectively. We will find that for the Fresnel polynomial \eqref{eq:FresnelPert} these are different functions. In case the WPE coefficients $\mathcal{K}^{abcd}$ vanish, i.e.\ on the basis of metric electrodynamics, there exists a close relation between them, namely one is the square root of the other.

To be able to obtain the aforementioned Lagrangians, it is necessary that the determinant of the second derivative of the Fresnel polynomial with respect to the momentum covectors $k$ does not vanish, i.e., $\det(\partial_{k_a}\partial_{k_b}\mathcal{G}(k))\neq 0$ for all $k$ satisfying $\mathcal{G}(k)=0$, see \cite{Raetzel:2010je}. Thus, our subsequent results hold if
\begin{align}
\begin{split}
	\det(\partial_{k_a}\partial_{k_b}\mathcal{G}(k))
	&= - 96 \eta^{-1}(k,k)^2 \bigg( 8 \eta^{-1}(k,k)^2 - \sigma 4 \eta^{-1}(k,k) \big(4 \mathcal{K}(k,k) + \mathcal{K}^a{}_a \eta^{-1}(k,k)\big)\\
	&+ \sigma^2 \big( 8 \mathcal{K}(k,k)^2 + 8 \mathcal{J}(k,k,k,k) + 4 \eta^{-1}(k,k) k_{a}k_b (\mathcal{K}_{ac}\mathcal{K}^c{}_{b} - 3 \mathcal{J}^{c}{}_{cab})\\
	&- \eta^{-1}(k,k)^2 \mathcal{K}_{ab}\mathcal{K}^{ab} + \eta^{-1}(k,k) \mathcal{K}^a{}_{a} (10 \mathcal{K}(k,k) + \eta^{-1}(k,k) \mathcal{K}^b{}_{b}) \big)   \bigg) + O(\mathcal{K}^3)\neq 0\,.
\end{split}	
\end{align}
To second order in $\mathcal K$ this means that, $\eta^{-1}(k,k)=0$ should not imply $\mathcal{G}(k)=0$. If we would have considered the Fresnel polynomial \eqref{eq:FresnelPert} only to first order that would be the case however. Hence, in order to be able to describe the radar experiment in terms of configuration space wordlines at all it is necessary to consider the second order WPE perturbation of the Fresnel polynomial from metric electrodynamics.

\subsection{Propagation of light}\label{ssec:lighlagrange}

Applying the methods developed in \cite{Raetzel:2010je}, the massless/lightlike directions $\dot x$, corresponding to the lightlike wave covectors $k$ determined by the vanishing of the Fresnel polynomial \eqref{eq:FresnelPert}, are described by a Lagrange function called~$L^\#$.
It is defined via the Helmholtz action for trajectories of massless momenta
\begin{align}\label{eq:Helmholtz_action}
S[x,k,\lambda] = \int \dif\tau \big( k_a\dot x^a - \lambda \mathcal{G}(k)  \big),
\end{align}
which can be successively transformed into an action
\begin{align}
S[x,\mu] = \mu \int \dif\tau L^\#(x, \dot x),
\end{align}
where $L^\#(y)$ is the so called dual polynomial to $\mathcal{G}(k)$. The dual polynomial is a polynomial in $y$ defined via the relation that for $y^a = \partial_{k_a}\mathcal{G}(k)$ it satisfies $L(\partial_{k_a}\mathcal{G}(k)) = Q(k) \mathcal{G}(k)$ for some factor $Q(k)$. In particular this means that $L^\#(y) = 0$ for the velocities $y^a = \partial_{k_a}\mathcal{G}(k)$ which are obtained from null momenta $k$ satisfying $\mathcal{G}(k)=0$. 

To obtain the dual polynomial we first we solve the equations of motion obtained by variation with respect to $k$ for $k$ as function of the velocities $\dot x$. 

Variation of Eq.\ \eqref{eq:Helmholtz_action} with respect to $k$ yields
\begin{align}\label{eq:y(k)}
\begin{split}
	y_a(k) = \frac{\dot x_a(k)}{\lambda} = \partial_{k_a}\mathcal{G}(k) 
	&=  4 k_{a} \eta^{-1}(k,k) - 2 ( \eta^{-1}(k,k) k^{b} \mathcal{K}_{ab} +  k_{a} \mathcal{K}(k,k))\\
	&- 2 (k_{b} k_{c} k_{d} \mathcal{J}_a{}^{bcd} - k^{b} \mathcal{K}_{ab} \mathcal{K}(k,k)) + O(\mathcal{K}^3),
\end{split}
\end{align}
which can be inverted to
\begin{align}
\begin{split}
	k_a(y) 
	&= \frac{y_{a}}{2^{2/3} \eta(y,y)^{1/3}} + \frac{1}{2^\frac{2}{3}} \bigg( \frac{\mathcal{K}_{ab}y^{b}}{2 \eta(y,y)^{1/3}} - \frac{\mathcal{K}(y^\flat,y^\flat) y_{a}}{6 \eta(y,y)^{4/3}} \bigg)  \\
	&+ \bigg[ y_a\bigg(\frac{2^{1/3} \mathcal{K}(y^\flat,y^\flat)^2}{9 \eta(y,y)^{7/3}} - \frac{\mathcal{K}_{bc}\mathcal{K}^{c}{}_{d} y^{b} y^{d}}{12\ 2^{2/3} \eta(y,y)^{4/3}} - \frac{\mathcal{J}(k,k,k,k)}{3\ 2^{2/3} \eta(y,y)^{7/3}} \bigg)  \\
	& + \frac{\mathcal{J}_{abcd} y^{b} y^{c} y^{d}}{2\ 2^{2/3} \eta(y,y)^{4/3}} - \frac{\mathcal{K}_{ab}y^b \mathcal{K}(y^\flat,y^\flat)}{3\ 2^{2/3} \eta(y,y)^{4/3}} + \frac{\mathcal{K}_{ac}\mathcal{K}^{c}{}_{d} y^{d}}{4\ 2^{2/3} \eta(y,y)^{1/3}}\bigg] + O(\mathcal{K}^3).
\end{split}	
\end{align}
We introduced the musical-isomorphism notation $y^\flat = \eta(y,\cdot)$ for a vector mapped to a covector by pulling an index with the inverse Minkowski metric.

Next, we find the non-polynomial Lagrangian $\tilde L$, which satisfies $\tilde L(\partial_{k_a}\mathcal{G}(k)) = \mathcal{G}(k)$
\begin{align}
\begin{split}
	\tilde L (y) 
	&= \mathcal{G}(k(y)) \\
	&= \frac{1}{2^\frac{8}{3} \eta(y,y)^{\tfrac{4}{3}}}(\eta(y,y)^2 + \frac{1}{3} \mathcal{K}(y^\flat,y^\flat) \eta(y,y)  \\
	&+ \frac{1}{18}\bigg(3 \mathcal{J}(y^\flat,y^\flat,y^\flat,y^\flat) - 2 \mathcal{K}(y^\flat,y^\flat)^2 + 3 \mathcal{K}_{ac} \mathcal{K}^{c}{}_{b} y^{a} y^{b} \eta(y,y)\bigg)
	+ O(\mathcal{K}^3)).
\end{split}
\end{align}
The desired dual polynomial $L^\#$ is not given by $2^\frac{8}{3} \eta(y,y)^{\tfrac{4}{3}} \tilde L$ since this polynomial does not satisfy the defining equation $L^\#(\partial_{k_a}\mathcal{G}(k)) = Q(k) \mathcal{G}(k)$. By simple modification of the coefficients in front of the different terms appearing in this polynomial, we find, however, easily that the desired dual polynomial is given by
\begin{align}\label{eq:Lsharp}
	L^\#(y) = \eta(y,y)^2 + 3 \mathcal{K}(y^\flat,y^\flat) \eta(y,y) + \frac{1}{2}(3 \mathcal{J}(y^\flat,y^\flat,y^\flat,y^\flat) + 4 \mathcal{K}(y^\flat,y^\flat)^2 + 5 \mathcal{K}_{ac} \mathcal{K}^{c}{}_{b} y^{a} y^{b} \eta(y,y))\,.
\end{align}
Using \eqref{eq:y(k)} one can verify by direct calculation that $L^\#(y(k)) = Q(k) \mathcal{G}(k)= (4 \eta^{-1}(k,k))^4 \mathcal{G}(k)$ as desired.

To determine the directions $N$ along which light propagates, we make the ansatz $N = N_0 + N_{1\pm}$, where $N_0$ is the leading order and $N_1$ of order $\mathcal{K}$. Evaluating the null-condition $L^\#(N) = 0 + O(\mathcal{K}^3)$ order by order the null vectors have to satisfy 
\begin{align}\label{eq:lightlike}
	\eta(N_0,N_0) = 0,\quad \eta(N_0,N_{1_\pm}) = - \frac{3}{4}\mathcal{K}(N_0^\flat, N_0^\flat) \pm \frac{1}{4}\sqrt{\mathcal{K}(N_0^\flat, N_0^\flat)^2 - 6 \mathcal{J}(N_0^\flat,N_0^\flat,N_0^\flat,N_0^\flat)}.
\end{align}
Thus, there exist two different sets of lightlike directions predicted by WPE. The effect we encounter here is nothing but birefringence, where different polarizations of light, here distinguished by different lightlike vectors, propagate in different null surfaces. This effect is unavoidable as soon as one leaves the realm of metric electrodynamics \cite{Lammerzahl:2004ww}.
In order to obtain well-defined light propagation one necessary requirement on the perturbation parameters $\mathcal{K}^{abcd}$~is 
\begin{align}
\mathcal{K}(N_0^\flat, N_0^\flat)^2 - 6 \mathcal{J}(N_0^\flat,N_0^\flat,N_0^\flat,N_0^\flat)\geq 0.
\end{align}

The calculation above shows again explicitly that it was necessary to go to second order in $\mathcal{K}^{abcd}$ in $G(k)$ to capture effects of the perturbation of the propagation of light, since this second order in \eqref{eq:Lsharp} enters the perturbation of the light direction $N_1$.

\subsection{Propagation of massive particles and observers}
To determine the Lagrange function $L^*$, which determines the motion of observers, i.e. particles with mass $m>0$, consistent with the theory of electrodynamics in consideration, we follow again the procedure introduced in \cite{Raetzel:2010je} and apply the action
\begin{align}
	S[x,k,\lambda] = \int \dif\tau \bigg( k_a\dot x^a - \lambda \ln\big(\mathcal{G}(\tfrac{k}{m})\big)  \bigg).
\end{align}
Variation with respect to $\lambda$ enforces the dispersion relation $\mathcal{G}(k) = m^4$. Variation with respect to $x$ yields the equation of motion for the time evolution of $k$ and variation with respect to $k$ gives the equation, which relates massive covectors to its corresponding velocities, where a first order calculation suffices here,
\begin{align}\label{eq:dotx(k)}
	y^a = \frac{\dot x^a}{\lambda} = \frac{1}{4} \frac{\bar{\partial}^a \mathcal{G}(k)}{\mathcal{G}(k)} =  \frac{k^a}{\eta^{-1}(k,k)} + \frac{1}{2} \frac{k^a \mathcal{K}(k,k)}{\eta^{-1}(k,k)^2} - \frac{1}{2} \frac{ \mathcal{K}^{ab}k_b}{\eta^{-1}(k,k)} + O(\mathcal{K}^2).
\end{align}
Observe that the right hand side of \eqref{eq:dotx(k)} is homogeneous of degree $-1$ with respect to $k$. Thus, $k_a$ must be homogeneous of degree $-1$ with respect to $\dot x$.
Inserting $k(\tfrac{\dot x}{\lambda})$ in the action to obtain $S[\dot x, \lambda]$ and performing variation with respect to $\lambda$ yields the dispersion relation in the form
\begin{align}
	\ln\bigg(\frac{\lambda^4}{m^4}\bigg) = \ln\bigg(\mathcal{G}(k(\dot x))\bigg),
\end{align}
which we can now solve for $\lambda(\dot x)$
\begin{align}
	\lambda = \frac{m}{\mathcal{G}(k(\dot x))^\frac{1}{4}}.
\end{align}
Due to the homogeneity of $k$ and $\mathcal{G}$, this implies $\mathcal{G}(k(\tfrac{\dot x}{\lambda}))=m^4$. Inserting such a solution $k_a(\frac{\dot x}{\lambda})$, the solution for $\lambda$ and the dispersion relation $\mathcal{G}(k) = m^4$ into the action yields
\begin{align}
	S[x] = m \int \dif\tau \frac{1}{\mathcal{G}(k(\dot x))^\frac{1}{4}}.
\end{align}
Note that the same reasoning can be applied for general r-homogeneous functions $\mathcal{G}(k)$ whose level sets define dispersion relations. In general then the action becomes
\begin{align}
S[x] =  m \int \dif\tau \frac{1}{\mathcal{G}(k(\dot x))^\frac{1}{r}}.
\end{align}

Returning to the Fresnel polynomial of WPE by solving Eq.\ \eqref{eq:dotx(k)} for $k_a(\frac{\dot x}{\lambda})$ explicitly, we find to first order in $\mathcal{K}$
\begin{align}\label{eq:k(dotx)}
	k^a(y) = \eta^{ab}k_b(y) = \frac{y^a}{\eta(y,y)} + \frac{1}{2} \bigg(\frac{\mathcal{K}^{ab}y_b}{\eta(y,y)} - \frac{y^a \mathcal{K}(y^\flat,y^\flat)}{\eta(y,y)^2}\bigg) + O(\mathcal{K}^2),
\end{align}
which implies
\begin{align}\label{eq:G(k(dotx))}
	\mathcal{G}(k(y)) 
	&= \frac{1}{\eta(y,y)^2} -  \frac{\mathcal{K}(y^\flat,y^\flat)}{\eta(y, y)^3} + O(\mathcal{K}^2).
\end{align}
Following the general argument presented above, we only need to calculate
\begin{align}
 \frac{1}{\mathcal{G}(k(\dot x))^{\frac{1}{4}}}
	&= \sqrt{\eta(\dot x,\dot x)} \bigg(1+\frac{1}{4}\frac{\mathcal{K}(\dot x^\flat, \dot x^\flat) }{\eta(\dot x, \dot x)}\bigg) + O(\mathcal{K}^2) 
\end{align}
to obtain the action for massive particles
\begin{align}\label{eq:Lstar}
	S[x] =  m \int \dif\tau \left[\sqrt{\eta(\dot x,\dot x)} \bigg(1+\frac{1}{4}\frac{\mathcal{K}(\dot x^\flat, \dot x^\flat) }{\eta(\dot x, \dot x)}\bigg) + O(\mathcal{K}^2)\right].
\end{align}
This defines the desired Lagrangian $L^*(\dot x) = \sqrt{\eta(\dot x,\dot x)} \big(1+\tfrac{1}{4}\tfrac{\mathcal{K}(\dot x^\flat, \dot x^\flat) }{\eta(\dot x, \dot x)}\big) + O(\mathcal{K}^2)$.
It determines the motion of massive particles and an observer's measurement of time. Note that since $L^*$ is homogeneous of degree one with respect to its argument it is a Finsler function.
Thus, WPE, cf.\ \eqref{eq:edynsme}, which contains the SME electrodynamics, predicts the motion of massive particles on Finslerian geodesics. To the knowledge of the authors, this connection between the SME and Finsler geometry has not been pointed out earlier. Further connections between the SME and Finsler geometry have been made in the literature, see e.g.\ \cite{Kostelecky:2011qz,Kostelecky:2012ac,Foster:2015yta}.

In contrast to the massless particles case derived in Sec.\ \ref{ssec:lighlagrange}, the first order expression of $L^*$ suffices for all upcoming applications.

\section{An observer's measurement of time and spatial length}\label{sec:timelength}
In the previous section, we derived the Lagrangians that describe the relativistic motion of light and observers determined by WPE:
\begin{itemize}
	\item  $L^\#$, defined in Eq.\ \eqref{eq:Lsharp}, determines the motion of light, i.e. the motion along curves whose tangent vector $N$ satisfies $L^\#(N) =0$. 
	\item  $L^*$, read-off from Eq.\ \eqref{eq:Lstar}, determines the motion and proper time of observers, i.e. the motion along curves whose tangent vectors $U$ satisfies $L^*(U)=1$.\footnote{A mathematical more precise identification of observer directions in terms of the hyperbolicity properties of $\mathcal{G}$ and $L^\#$ was developed in \cite[Sec.\ 5]{Raetzel:2010je} and summarized in \cite[Sec.\ 2.2]{Fewster:2017mtt}. Observer directions are those which lie inside a hyperbolicity cone of $L^\#$.}
\end{itemize}
The Lagrangian $L^*$ defines an observer's measurement of time and its equal time surfaces. It thus determines time dilation effects as we will see in more detail in Sec.\ \ref{ssec:tdil}. In contrast, an observer's measurement of spatial length in terms of a radar experiment is influenced by both Lagrangians $L^*$ and $L^\#$. The resulting radar length determines length contractions, which we derive in Sec.\ \ref{ssec:lcon}.

\subsection{The clock postulate and equal time surfaces}\label{ssec:clock}
One of the fundamental postulates special and general relativity are based on is the clock postulate: The time that an observer measures between two events is the length of its worldline $\gamma$ connecting these events. In special and general relativity, this postulate is realized by specifying the metric length of curves, which is the zeroth order of $L^*$ in \eqref{eq:Lstar}, to be the relevant length measure. 

We can use the clock postulate to define the time measurement of observers in non-metric background geometries. The Lagrangian $L^*$, which governs the motion of massive particles defines a length functional for observer curves~$\gamma$ with an arbitrary parametrization $\tau$ and respective tangent $\dot \gamma=\frac{d\gamma}{d\tau}$. The length of the worldline between two events $p_1 = \gamma(\tau_1)$ and $p_2 = \gamma(\tau_2)$ is given by
\begin{align}\label{eq:time}
	T[p_1, p_2,\gamma] = \int_{\tau_1}^{\tau_2} \dif\tau\ L^*(\dot\gamma)\,.
\end{align}
It is interpreted as time the observer measures between them.
Since $L^*$ is $1$-homogeneous in its vector argument the length functional is parametrization invariant and there exits a distinguished parametrization of the curve, its arc length parametrization
\begin{align}
	T[p_1, p_2,\gamma] = \int_{\tau_1}^{\tau_2} \dif\tau\ L^*(\dot\gamma) = \int_{t(\tau_1)}^{t(\tau_2)} \dif t\ L^*(\gamma') = t(\tau_2) - t(\tau_1)\,.
\end{align}
This parametrization is an observer's proper time and the respective tangent vector denoted by $U(t) = \gamma'(t) = \frac{d\gamma(t)}{dt}$ satisfies $L^*(U(t)) = 1$. Thus, the demand of the normalization of the observer's tangent with respect to $L^*$ fixes the parametrization of the observer curve to arc-length parametrization.

The equal time surfaces $\Sigma_t$ of an observer are then mathematically characterized by the vectors $X$, which are annihilated by the momentum covector $k(U(t))$ corresponding to the observers tangent $U(t)$, see \cite[Sec.\ 7]{Raetzel:2010je}. More precisely the set of \emph{spatial vectors}
\begin{align}\label{eq:spatialset}
T_{\gamma(t)}\Sigma_t = S_{\rm ET}(U(t)) \defn \big\{\ X \in T_{\gamma(t)} M\ |\ k_a(U(t))X^a = 0\ \big\}\,,
\end{align}
is the tangent space to the equal time surface at the observer position. With help of Eq.\ \eqref{eq:k(dotx)} and the normalization condition
\begin{align}\label{eq:ObsNorm1}
L^*(U) = \sqrt{\eta(U,U)} \bigg(1+\epsilon \frac{1}{4} \frac{\mathcal{K}(U^\flat, U^\flat)}{\eta(U,U)}\bigg) = 1 \Rightarrow \eta(U,U) = 1 - \epsilon \frac{1}{2}\mathcal{K}(U^\flat, U^\flat)\,,
\end{align}
the defining relation can be expressed as
\begin{align}\label{eq:spatial}
	k_a(U(t))X^a = \eta(U(t),X) + \frac{1}{2}\mathcal{K}(U(t)^\flat,X^\flat)= 0\,.
\end{align}
In metric spacetime geometry, the spatial vectors are given by the orthocomplement of $U(t)$ with respect to the spacetime metric. Here, the modification to that condition induced by the causal structure derived from WPE \eqref{eq:edynsme} becomes explicitly visible.

So far we showed how WPE fixes the propagation of light as well as an observer's measurement of time and the observer's spatial directions or equal time surfaces. Next, we use a radar experiment to determine an observers notion of spatial length.

\subsection{Radar length}\label{ssec:radarlength}
\begin{figure}
	\centering
	\begin{subfigure}[b]{0.3\textwidth}
		\includegraphics[width=0.5\textwidth]{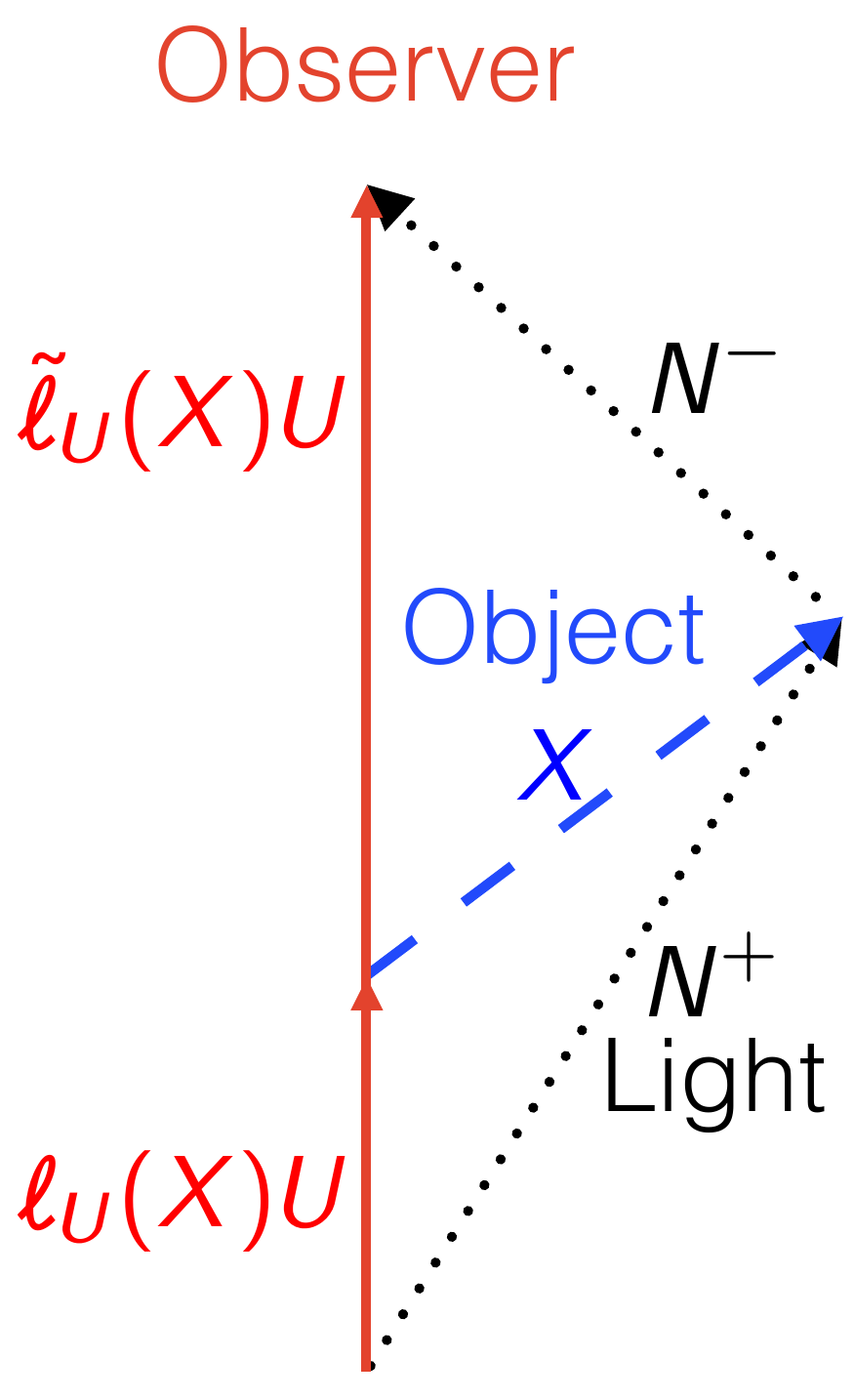}
		\caption{The general radar experiment}
		\label{fig:general}
	\end{subfigure}
	~ 
	\begin{subfigure}[b]{0.3\textwidth}
		\includegraphics[width=0.5\textwidth]{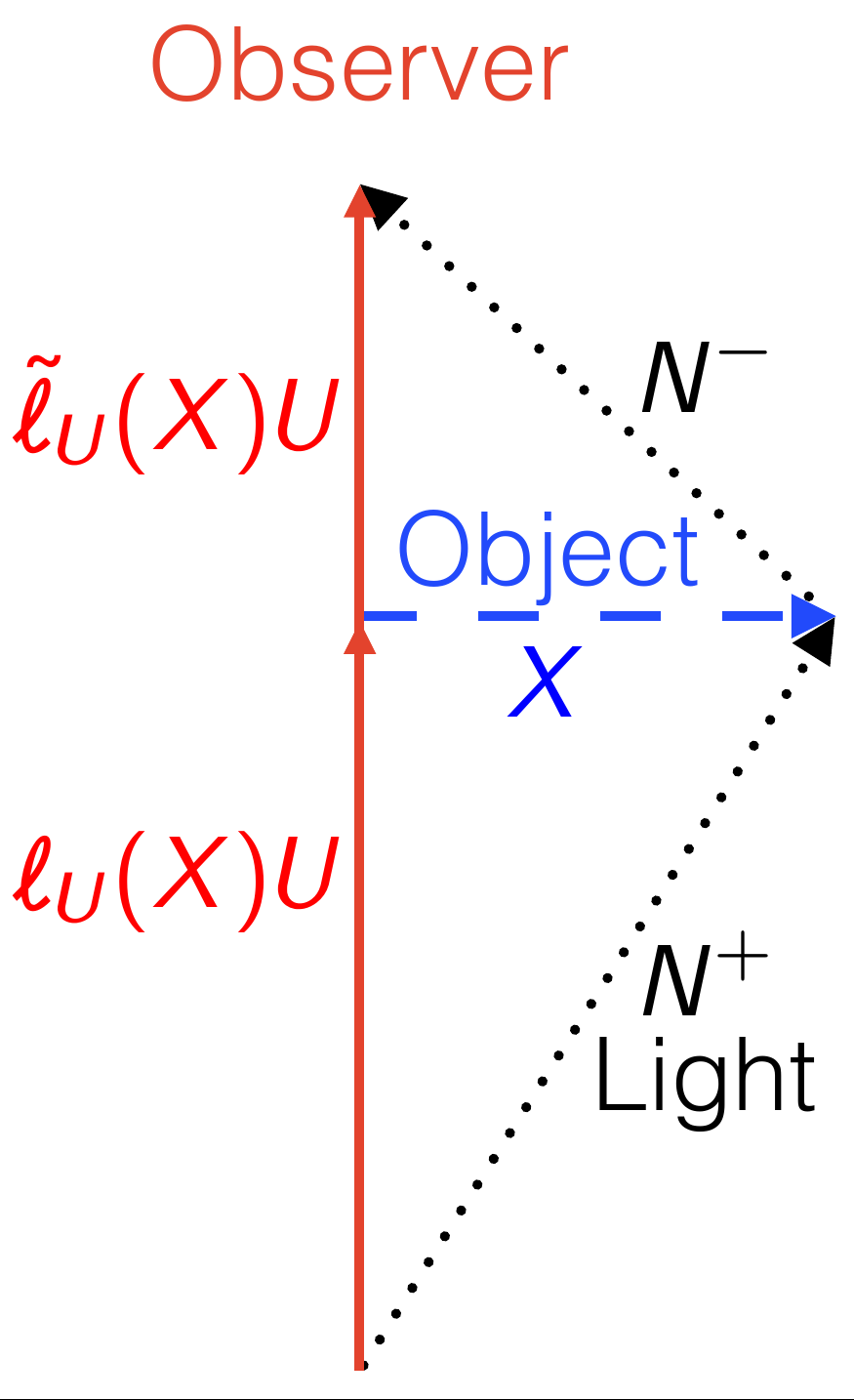}
		\caption{$X$ spatial with respect to $U$}
		\label{fig:ets}
	\end{subfigure}
	~ 
	\begin{subfigure}[b]{0.3\textwidth}
		\includegraphics[width=0.5\textwidth]{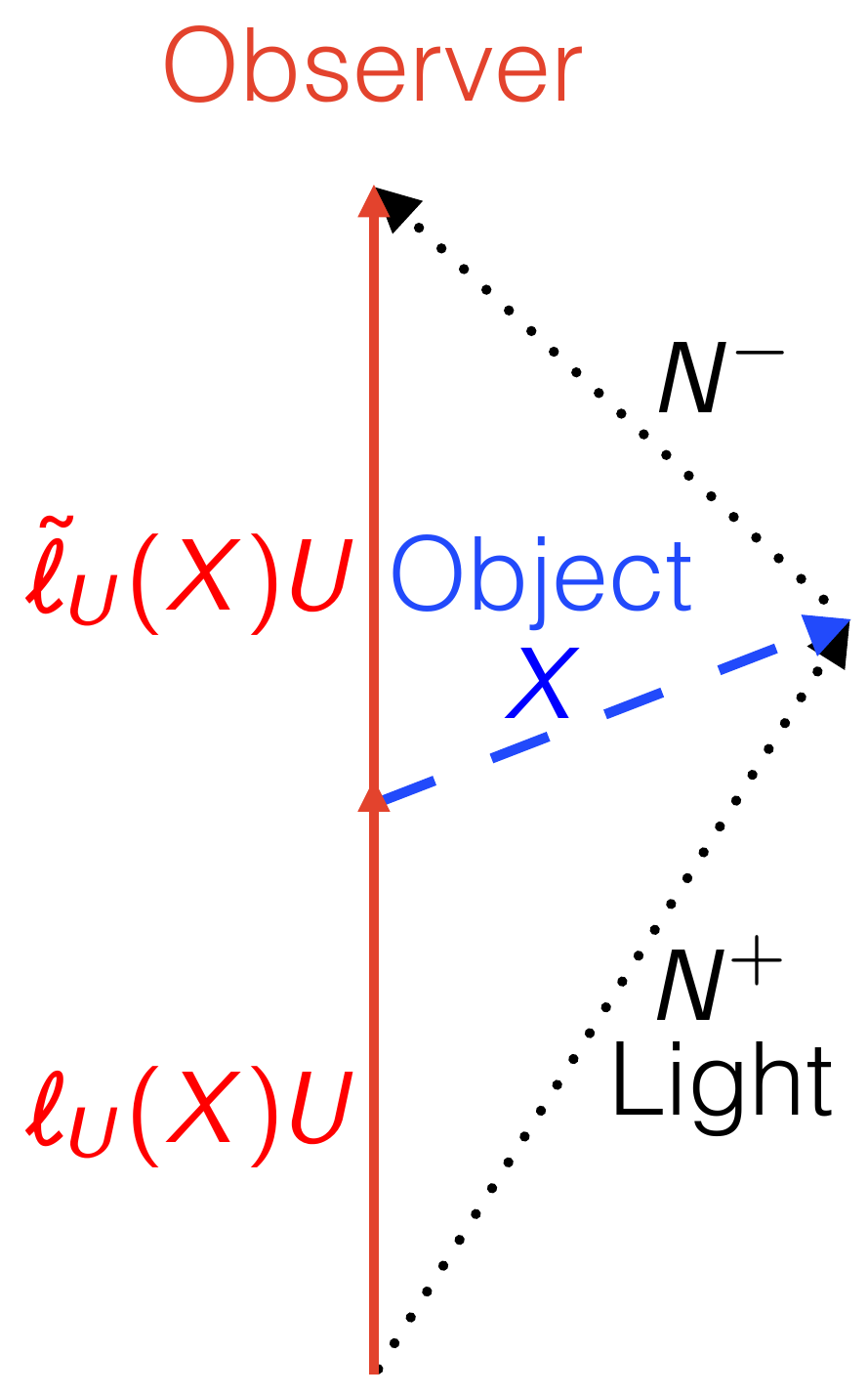}
		\caption{$\ell_U(X)=\tilde\ell_U(X)$}
		\label{fig:symmetric}
	\end{subfigure}
	\caption{Sketch of a radar experiments: An observer on its worldline (solid) with tangent $U$ emits light (dotted) $(N^+)$ along an object $X$ (dashed) which propagates back $(N^-)$ to the observer.}
	\label{fig:radarsk}
\end{figure}

Suppose $X$ is an object carried by an observer on a worldline $\gamma$ with $L^*$ normalized tangent $U$ who is located at one endpoint of $X$. At the other end of $X$ a mirror shall be attached. The observer on $\gamma$ obtains the radar length $\mathcal{L}_{U}(X)$ of $X$ by measuring the proper time a light pulse needs to travel from the observer to the mirror and back as sketched in Fig.\ \ref{fig:radarsk}. For the analysis of the radar experiment we assume that $U$ is constant during the time the experiment is performed and that $X$ is modelled by a constant vector, i.e., we assume that the observer is inertial and that there are no internal changes of the object induced by WPE over the travel time of the light pulse.


On the basis of a general Finslerian clock, the radar length has been described in \cite{Pfeifer:2014yua}. The difference in the analysis we perform here is that we relax some assumptions. First, we will not assume anything about the constancy or non-constancy of the spatial speed of light an observer measures, this quantity is determined by the underlying theory of electrodynamics. Second, we do not set $L^*=L^\#$, since this is in general not the case as we have seen in Sec.~\ref{sec:PartProp} and third we do not assume that the radar experiment is symmetric, since this setup does in general not measure equal time distances, as will become clear in Sec.\ \ref{ssec:radarsymmetric}.

The key relation in the infinitesimal or flat spacetime description of the radar experiment is that the sum and the difference of certain multiples of an observer's normalized tangent $U$ and the direction $X$ must yield a lightlike direction, cf.\ Fig.\ \ref{fig:radarsk}. More precisely,
\begin{align}\label{eq:nullvectors}
	N^+ = \ell_U(X) U + X,\text{ and } N^- = \tilde \ell_U(X) U - X
\end{align}
have to satisfy
\begin{align}\label{eq:radarcond}
	L^\#(N^+) = 0 = L^\#( N^-)\,,
\end{align}
and $U$ is normalized with respect to $L^*$ 
\begin{align}\label{eq:Unorm}
	L^*(U) = 1.
\end{align}
The null conditions \eqref{eq:radarcond} determine the factors $\ell_U(X)$ and $\tilde \ell_U(X)$ and their sum gives the radar length
\begin{align}\label{eq:radartime}
	\mathcal{L}_U(X)=\ell_U(X)+\tilde \ell_U(X).
\end{align}
Observe that the units of the radar length here are units of time. Since we do not assume a constant speed of light there is no natural conversion factor between units of length and units of time available.
We will derive the explicit expressions for  $\ell_U(X)$ and $\tilde \ell_U(X)$ and discuss different choices of $X$ along which a radar experiment may be performed.

As sketched in Fig.\ \ref{fig:radarsk}, the same radar time measurement belongs to different choices of $X$, i.e. the radar experiment does not determine $X$. The description of the radar experiment depends on additional conditions which prescribe the relation between the $X$ and $U$. In fact, the ambiguity is in choosing the starting point of $X$ along the worldline of the observer. One well-motivated way, depicted in Fig.\ \ref{fig:ets}, would be to consider only those $X$ that are spatial with respect to $U$, i.e., that are in its equal time surface and satisfy \eqref{eq:spatial}, see Sec.\ \ref{ssec:clock}. Another possibility, depicted in Fig.\ \ref{fig:symmetric}, is to choose $X$ such that the radar experiment is symmetric, i.e.\ $\ell_U(X)=\tilde\ell_U(X)$. On the basis of metric electrodynamics both of these choices are identical, however for general WPE this is no longer the case.

\subsubsection{The setup and the zeroth order result}
To present the results in a structured way, we consider a general light Lagrangian $L^\#$ as a power series in a perturbation parameter $\epsilon$, which parametrizes the deviation from metric light propagation. It can be used for various applications depending on the source of the deviation. Here, it corresponds to the powers of the perturbation $\mathcal{K}$ of metric electrodynamics discussed in Sec.\ \ref{sec:PartProp}. Thus, we use the light Lagrangian
\begin{align}\label{eq:Lstarsecorder}
	L^\#(y) = G(y,y,y,y) = G_{abcd}y^ay^by^cy^d = \eta(y,y)^2 + \epsilon \eta(y,y)h_1(y,y) + \epsilon^2 h_2(y,y,y,y)\,,
\end{align}
where the perturbation tensors $h_1$ and $h_2$ can be read off Eq.\ \eqref{eq:Lsharp} and $G$ is the totally symmetric light Lagrangian tensor.
A short calculation determines its components in a coordinate basis
\begin{align}\label{eq:componentsG}
	G_{abcd} 
	&= \frac{1}{4!}\partial_{y^a}\partial_{y^b}\partial_{y^c}\partial_{y^d}L^\# = \frac{1}{3} \big(\eta_{ab} \eta_{cd} + \eta_{ac} \eta_{bd} + \eta_{ad} \eta_{bc}\big)\nonumber\\
	&+ \frac{\epsilon}{6}\big(\eta_{ad}h_1{}_{bc}+\eta_{ac}h_1{}_{bd}+\eta_{ab}h_1{}_{cd}+\eta_{bc}h_1{}_{ad}+\eta_{bd}h_1{}_{ac}+\eta_{cd}h_1{}_{ab}\big) +\epsilon^2 h_2{}_{abcd}\,,
\end{align}
with
\begin{align}
	h_{1ab}
	= 3\mathcal{K}_{ab}\,,\quad
	h_{2abcd}
	&=\frac{1}{2}\bigg(3 \mathcal{K}{}_{(a}{}^{r}{}_{b}{}^{s}\mathcal{K}_{c|s|d)r} + 4 \mathcal{K}_{(ab}\mathcal{K}_{cd)} + 5 \eta_{(ab}\mathcal{K}_{c}{}^r\mathcal{K}_{d)r}\bigg)\,.
\end{align}
In the same language, the normalization of the observer tangent $U$ reads, c.f.\ Eq.\ \eqref{eq:ObsNorm1},
\begin{align}\label{eq:unorm}
	L^*(U)  = 1 \Rightarrow \eta(U,U) = 1 - \epsilon \frac{1}{6}h_{1}(U,U)\,.
\end{align}

To solve the null conditions \eqref{eq:radarcond} to first order in $\epsilon$, we write
\begin{subequations}
\begin{align}
	\ell_U(X) &= \ell^{(0)}_U(X) + \epsilon\ \ell^{(1)}_U(X)\,,\\
	\tilde \ell_U(X) &= \tilde \ell^{(0)}_U(X) + \epsilon\ \tilde \ell^{(1)}_U(X)\,.
\end{align}
\end{subequations}
In addition, we decompose the scalar product between $U$ and $X$:
\begin{align}\label{eq:XUpert}
	\eta(X,U) = f^{(0)}(X,U) + \epsilon\ f^{(1)}(X,U)\,.
\end{align}
Specifying the functions $f^{(0)}(X,U)$ and $f^{(1)}(X,U)$ fixes relations between $X$ and $U$, i.e. singles out special direction along which the radar experiment can be performed. Some specific choices will discussed below.

To zeroth order the Eqs.\ \eqref{eq:radarcond} are solved by
\begin{subequations}\label{eq:zeroorderradarlength}
\begin{align}
	\ell^{(0)}_U(X) &= - f^{(0)}(X,U) + \sqrt{f^{(0)}(X,U)^2 - \eta(X,X)}\\
	\tilde \ell^{(0)}_U(X) &= f^{(0)}(X,U) + \sqrt{f^{(0)}(X,U)^2 - \eta(X,X)}
\end{align}
\end{subequations}
whose sum is 
\begin{align}
\mathcal{L}_U(X) = \ell^{(0)}_U(X)+\tilde \ell^{(0)}_U(X) = 2  \sqrt{f^{(0)}(X,U)^2 - \eta(X,X)}\,.
\end{align}
Hence, for an observer's spatial distances satisfying $f^{(0)}(X,U) = 0$, i.e.\ those vectors orthogonal to $U$ with respect to the Minkowski metric $\eta$, see Eq.\ \eqref{eq:spatial}, the usual spatial length expression known from special relativity 
\begin{align}
\ell_U(X) = \tilde \ell_U(X) = \sqrt{-\eta(X,X)}
\end{align}
is obtained, derived here from metric electrodynamics.

Important to mention is that this zero order solutions also solve the first order of the null condition Eq.~\eqref{eq:radarcond}. This demonstrates once more explicitly why we had to go to second order in perturbation theory when we derived the geometric optics limit, c.f.\ Eq.\ \eqref{eq:FresnelPert}.

\subsubsection{The first order weak premetric perturbation}
The second order of the null conditions \eqref{eq:radarcond} determine the first order corrections of $\ell_U(X)$ and $\tilde \ell_U(X)$ from WPE.\footnote{The calculation was done with help of the computer algebra extension xAct for mathematica.} We find the four solutions
\begin{subequations}\label{eq:firstorderradarlength}
\begin{align}
	\ell^{(1)}_U(X)_\sigma &= \frac{1}{\left(\ell^{(0)}_U(X) + f^{(0)}(X,U)\right)}\left(A(X,U) + \sigma \frac{1}{4}\sqrt{B(X,U)}\right)\\
	\tilde \ell^{(1)}_U(X)_{\tilde \sigma} &= \frac{1}{\left(\tilde \ell^{(0)}_U(X) - f^{(0)}(X,U)\right)}\left(\tilde A(X,U) + \tilde \sigma \frac{1}{4}\sqrt{\tilde B(X,U)}\right)\,,
\end{align}
\end{subequations}
where $\sigma$ and $\tilde \sigma$ can assume the values $1$ or $-1$ and  we introduced the abbreviations
\begin{subequations}\label{eq:A}
\begin{align}
	A(X,U) &= - \frac{1}{4}h_1(X,X) - \frac{1}{6}\ell_U^{(0)}(X) \big( 3 \ell_U^{(0)}(X) h_1(U,U) + h_1(X,U) + 6 f^{(1)}(X,U) \big)\,,\\
	\tilde A(X,U) &= - \frac{1}{4}h_1(X,X) - \frac{1}{6}\tilde \ell_U^{(0)}(X) \big(3 \tilde \ell_U^{(0)}(X) h_1(U,U) - h_1(X,U) - 6 f^{(1)}(X,U) \big) \big)\,,
\end{align}
\end{subequations}
as well as
\begin{subequations}
\begin{align}
\begin{split}
	B(X,U) 
	&= h_1(X,X)^2 - 4 h_2(X,X,X,X) + 4 \ell_U^{(0)}(X) \big( h_1(X,X)h_1(X,U) - 4 h_2(X,X,X,U) \big)\\
	&+ 2 \ell_U^{(0)}(X)^2 \big( h_1(X,X)h_1(U,U) + 2  h_1(X,U)^2  - 12 h_2(X,X,U,U)  \big) \\
	&+   4 \ell_U^{(0)}(X)^3 \big( h_1(X,U)h_1(U,U) - 4 h_2(X,U,U,U) \big) + \big( h_1(U,U)^2 - 4 h_2(U,U,U,U) \big) \ell_U^{(0)}(X)^4\,,
\end{split}\\
\begin{split}
	\tilde B(X,U) 
	&= h_1(X,X)^2 - 4 h_2(X,X,X,X) - 4 \tilde \ell_U^{(0)}(X) \big( h_1(X,X)h_1(X,U) - 4 h_2(X,X,X,U) \big)\\
	&+ 2 \tilde\ell_U^{(0)}(X)^2 \big( h_1(X,X)h_1(U,U) + 2  h_1(X,U)^2  - 12 h_2(X,X,U,U)  \big) \\
	&-   4 \tilde\ell_U^{(0)}(X)^3 \big( h_1(X,U)h_1(U,U) - 4 h_2(X,U,U,U) \big) + \big( h_1(U,U)^2 - 4 h_2(U,U,U,U) \big) \tilde\ell_U^{(0)}(X)^4\,.
\end{split}
\end{align}
\end{subequations}
Moreover we used the normalization formula \eqref{eq:unorm} to eliminate $\eta(U,U)$. As the zeroth order solutions, the expressions obtained are $1$-homogeneous in $X$ what gives them indeed a scaling behavior of a length.

Observe that two effects influence these first order results for the radar length:
\begin{itemize}
	\item The modified propagation of light which enters through the second order of the radar conditions defined by the light Lagrangian $L^\#$ \eqref{eq:radarcond}.
	\item The modified observer proper time normalization which manifests itself in the normalization of $U$ with respect to the Lagrangian $L^*$ \eqref{eq:unorm}.
\end{itemize}
Both needed to be taken into account when studying observable effects from perturbations of metric electrodynamics.

\subsubsection{Interpretation}
Following Eqs.\ \eqref{eq:radartime} and \eqref{eq:firstorderradarlength}, there are four possible radar lengths an observer can measure depending on the setup of the experiment. 
In general, the light Lagrangian \eqref{eq:Lstarsecorder} predicts the propagation of light along trajectories with tangents $N$ which span two different null surfaces characterized by 
\begin{align}\label{eq:nullsurface}
\eta(N,N)_{\pm} = -\frac{1}{2}\epsilon h_1(N,N) \pm \epsilon \frac{1}{2} \sqrt{h_1(N,N)^2-4h_2(N,N,N,N)}\,.
\end{align}
We call the null trajectories with tangents satisfying the condition $\eta(N,N)_\pm$ elements of the null-surface $\mathcal{N}^\pm$. The outcome of the radar experiment depends on the kind of light propagating in its different parts, cf.\ Fig.\ \ref{fig:radarsk}.

The null vectors we found from the radar conditions have the following properties
\begin{itemize}
	\item $N^+_I = \big( \ell^{(0)}_U(X) + \epsilon\ \ell^{(1)}_U(X)_+ \big) U + X \in \mathcal{N}^+$,
	\item $N^+_{II} = \big( \ell^{(0)}_U(X) + \epsilon\ \ell^{(1)}_U(X)_- \big) U + X\in \mathcal{N}^-$,
	\item $N^-_I = \big( \tilde \ell^{(0)}_U(X) + \epsilon\ \tilde \ell^{(1)}_U(X)_+ \big) U - X\in \mathcal{N}^+$,
	\item $N^-_{II} = \big( \tilde \ell^{(0)}_U(X) + \epsilon\ \tilde \ell^{(1)}_U(X)_- \big) U - X\in \mathcal{N}^-$,
\end{itemize}
and so four different combinations $A=(N^+_I,N^-_I)$, $B=(N^+_{II},N^-_I)$, $C=(N^+_I,N^-_{II})$ and $D=(N^+_{II},N^-_{II})$ are possible to realize a radar experiment. These four choices differ in the chosen polarization of the light used. Hence, in interferometry experiments or in frequency comparison experiment based on cavities, in which basically the radar length of different distances is compared, it is important to specify with which kind of light the experiment was performed.

For all four possible radar experiment outcomes based on WPE the predicted radar lengths are
\begin{align}\label{eq:radar}
\begin{split}
	\mathcal{L}_U(X)_{\sigma \tilde \sigma} &= \ell_U(X)_{\sigma}+\tilde \ell_U(X)_{\tilde \sigma} \\ 
	&= 2 \sqrt{f^{(0)}(X,U)^2 - \eta(X,X)} + \epsilon \Big ( \ell^{(1)}_U(X)_\sigma + \tilde \ell^{(1)}_U(X)_{\tilde \sigma} \Big)  \\
	&= 2 \sqrt{f^{(0)}(X,U)^2 - \eta(X,X)}
	+ \frac{\epsilon}{4} \frac{ \sigma \sqrt{B(X,U)} + \tilde{\sigma} \sqrt{ \tilde B(X,U)}}{\sqrt{f^{(0)}(X,U)^2 - \eta(X,X)}} \\
	&+ \frac{\epsilon}{2} \frac{ 2 f^{(0)}(X,U) \Big(f^{(1)}(X,U) - \frac{1}{6}h_1(X,U) \Big) - 2 h_{1}(U,U) ( 2f^{(0)}(X,U)^2 - \eta(X,X)) -  h_1(X,X)  }{\sqrt{f^{(0)}(X,U)^2 - \eta(X,X)}},
\end{split}
\end{align}
where $	\mathcal{L}_U(X)_{++}$ is the travel time measured in the radar experiment $A$, $	\mathcal{L}_U(X)_{-+}$ in $B$, $	\mathcal{L}_U(X)_{+-}$ in $C$ and $	\mathcal{L}_U(X)_{--}$ in $D$.

The radar experiment based on WPE allows an observer to associate four different operationally accessible numbers, the radar radar lengths $\mathcal{L}_U(X)_{\sigma\tilde{\sigma}}$, to the vector $X$ which models an object here. Thus, the numbers $\mathcal{L}_U(X)_{\sigma\tilde{\sigma}}$ can also be interpreted as lengths an observer associates to vectors $X$, at least for all vectors for which these numbers are real. \footnote{If these radar length are really norms, or seminorms of vectors in the mathematical sense is an open questions and needs to be investigated in the future. They are $1$-homogeneous with respect to $X$. For being a proper mathematical (semi)norm a proof of satisfying the triangle inequality is missing and beyond the scope of this article.}

In special and general relativity, which correspond to the zeroth order solution we found, the different radar lengths are all identical. Moreover considering the length of spatial vectors, i.e. those satisfying $\eta(X,U)=f^{(0)}(X,U)=0$, is equivalent to a symmetric radar experiment setup $\ell^{(0)}_U(X)=\tilde \ell^{(0)}_U(X)$. For perturbations around this setup the radar signal measured is independent of $f^{(1)}$ due to the vanishing of the zeroth order $f^{(0)}$ and always of the form 
\begin{align}\label{eq:radarspatialsymm}
\begin{split}
\mathcal{L}_U(X)_{\sigma\tilde{\sigma}}
&= 2 \sqrt{- \eta(X,X)}\\
&+ \frac{\epsilon}{4} \frac{\ \bigg(\ \sigma \sqrt{B(X,U)}  + \tilde \sigma \sqrt{ \tilde B(X,U)}\ \bigg) \bigg|_{\ell^{(0)}_U(X) = \tilde \ell^{(0)}_U(X) = \sqrt{-\eta(X,X)}}}{\sqrt{- \eta(X,X)}} 
- \frac{\epsilon}{2} \frac{h_1(X,X) - 2 h_1(U,U)\eta(X,X)}{\sqrt{ - \eta(X,X)}}\,.
\end{split}
\end{align}

An interesting question we address now, is if there exist any weak premetric theories of electrodynamics for which the symmetric radar experiment $\ell_U(X)=\tilde \ell_U(X)$ measures spatial $X$ with respect to $U$ or if this is a unique feature of special and general relativity.

\subsubsection{Equal time surface and symmetry of the radar experiment}\label{ssec:radarsymmetric}
To zeroth order the radar experiment for spatial direction is symmetric, as previously discussed. Assuming this zeroth order setup the difference in the factors $\ell_U(X)$ and $\tilde \ell_U(X)$ derived from \eqref{eq:firstorderradarlength} and \eqref{eq:A} is
\begin{align}
	&\ell_U(X) - \tilde \ell_U(X) \nonumber \\
	&= \ell^{(1)}_U(X) - \tilde \ell^{(1)}_U(X) \nonumber \\
	&= -\frac{1}{3} h_1(X,U) - 2 f^{(1)}(X,U) + \frac{1}{4} \frac{1}{\ell^{(0)}_U(X)} \bigg( \sigma \sqrt{B(X,U)} - \tilde \sigma \sqrt{\tilde B(X,U)} \bigg) \bigg|_{\ell^{(0)}_U(X) = \tilde \ell^{(0)}_U(X) = \sqrt{-\eta(X,X)}}
\end{align}
The radar experiment is symmetric if and only if $\ell_U(X) - \tilde \ell_U(X)=0$ which thus implies
\begin{align}\label{eq:radarsymm}
f^{(1)}(X,U) = - \frac{1}{6} h_1(X,U) + \frac{1}{8} \frac{1}{\ell^{(0)}_U(X)} \bigg( \sigma \sqrt{B(X,U)} - \tilde \sigma \sqrt{\tilde B(X,U)} \bigg) \bigg|_{\ell^{(0)}_U(X) = \tilde \ell^{(0)}_U(X) = \sqrt{-\eta(X,X)}}\,.
\end{align}
The question is if the distances which are measured with a symmetric radar experiment, correspond to spatial distances for which $f^{(1)}(X,U) = - \frac{1}{6}h_1(X,U)$, see \eqref{eq:spatial}. In order that both conditions on $f^{(1)}$ can be satisfied simultaneously the perturbations tensors $h_1$ and $h_2$ of the light Lagrangian \eqref{eq:Lstarsecorder} must be related such that
\begin{align}\label{eq:ETSandSymmetric}
0 = \bigg( \sigma \sqrt{B(X,U)} - \tilde{\sigma} \sqrt{\tilde B(X,U)} \bigg) \bigg|_{\ell^{0}_U(X) =\tilde \ell^{(0)}_U(X)= \sqrt{-\eta(X,X)}}\,.
\end{align}
for all $U$ and $X\in S_{\rm ET}(U)$ for one combination of signs $\sigma,\tilde \sigma$. Thus, an observer can measure distances in its equal time surfaces with a symmetric radar experiment if and only if
\begin{align}
\sigma=\tilde \sigma\quad B(X,U) = \tilde B(X,U)\,.
\end{align}

Depending on the properties of the WEP perturbation parameters $\mathcal{K}^{abcd}$, this equation may either be satisfied for all observer tangents $U$ and their spatial directions $X$ or it might be satisfied only for certain observers $U$, who then perform a symmetric radar experiment along certain spatial directions $X$. An example, and maybe the only one, for the first case is $\mathcal{K}^{abcd}=0$, i.e.\, metric electrodynamics, while the example $\mathcal{K}^{abcd} = 4( Z^{[a}Y^{b]}V^{[c}W^{d]} + Z^{[c}Y^{d]}V^{[a}W^{b]})$, where $V,W,Y,Z$ are mutually orthogonal vector fields with respect to the Minkowski metric, represents an example of the second case, where \eqref{eq:ETSandSymmetric} is not identically satisfied. Nonetheless, assuming that $Y$ is the observer's direction, i.e.\ $U= Y$, Eq.\ \eqref{eq:ETSandSymmetric} is still satisfied for all of the observer's spatial directions $X$. It stands to reason that even in the general case, one can always find at least one observers $U$ that admits an spatial direction $X$ so that Eq.\ \eqref{eq:ETSandSymmetric} is satisfied.

\section{Classical relativistic observables}\label{sec:classrel}

The notion of time, equal time surfaces, spatial directions and radar length are the building blocks for classical relativistic observables like time dilation and length contractions, which we will briefly discuss subsequently. On the basis of metric electrodynamics, these notions are all connected to the norm of vectors measured with the spacetime metric and metric orthogonality of vectors. For WPE, these concepts manifest themselves in different mathematical expressions, as we just derived in the previous section. Detailed descriptions of the the Michelson–Morley, Ives-Stilwell and Kennedy-Thorndike experiment, which measure the relativistic effects and their deviation from predictions based on metric electrodynamics, on the basis of the notions introduced here are work in preparation.

\subsection{Time dilation}\label{ssec:tdil}
Consider two observers on worldlines $\gamma_{1}(t_1)$ and $\gamma_2(t_2)$, where each is parametrized with their proper time. The tangents $U_1 = \frac{d \gamma_1}{dt_1}$ and $U_2= \frac{d \gamma_2}{dt_2}$ are assumed to be constant, respectively, i.e.\ the observers are inertial and follow geodesics on the flat background manifold we are considering. According to the clock postulate, proper time parametrization implies $L^*(U_1)=L^*(U_2)=1$, see Sec.\ \ref{ssec:clock}. The observers meet at a certain point $p = \gamma_1(0)=\gamma_2(0)$, on which they synchronize their clocks to a fixed value, here chosen to be $t_1=t_2 = 0$.

After proper time $t_{1}$, the first observer decomposes the tangent of the second observer into a displacement vector $X_1 = t_1 V_1$  expressed in terms of the relative velocity $V_1$ of $\gamma_2$ relative to $\gamma_1$ and its own tangent
\begin{align}\label{eq:timedilationVdef}
t_{2} U_2= t_1 U_1 + X_1 = t_1 (U_1 + V_1)\,.
\end{align}
Note that we did not yet constrain ourselves to a specific way to construct the displacement vector and hence the relative velocity, except the restriction that $U_1$ and $U_2$ are both observer tangents.

The time $t_2$ is now easily obtained as function of $t_1, U_1$ and $V_1$ by taking the norm of the equation with $L^*$ using the normalization of $U_1$ and $U_2$, cf.\ Eq.\ \eqref{eq:unorm}, as well as the expansion of $\eta(V_1,U_1)$, c.f.\ Eq.\ \eqref{eq:XUpert} with $X$ set to $t_1 V_1$,\
\begin{align}\label{eq:time_dilation}
t_2 
&= L^*(t_1 U_1 + X_1) \nonumber \\
&=t_1 \sqrt{1 + 2 f^{(0)}(V_1,U_1) + \eta(V_1,V_1)}  \bigg(1 +  \epsilon\frac{2 ( h_1(V_1,U_1) + 6 f^{(1)}(V_1,U_1)) +  h_1(V_1,V_1)}{12(1 + 2 f^{(0)}(V_1,U_1) + \eta(V_1,V_1))}\bigg)\,.
\end{align}
The only additional assumption we used here is that $f^{(0)}(V_1,U_1)$ and $f^{(1)}(V_1,U_1)$ are homogeneous of degree one with respect to their first argument as they are for the two choices of interest we discuss now.

So far the separation vector $X_1$, respectively the relative velocity, could be chosen arbitrarily as long as it represents the difference between the two observer wordlines. Between all possible choices there are two classes of separations, maybe the most physical ones, which demonstrate the difference to the special relativistic case: 
\begin{itemize}
	\item The separations built from $V_1$ which are spatial with respect to $\gamma_1$, c.f. Eq. \eqref{eq:spatial}
	\begin{align}
	f^{(0)}(V_1,U_1)=0,\quad f^{(1)}(V_1,U_1)=-\frac{1}{6}h_1(V_1,U_1)\,.
	\end{align}
	For these, the time dilation becomes
	\begin{align}\label{eq:time_dilation2} 
	t_2 &=t_1\sqrt{1 + \eta(V_1,V_1)}  \bigg(1 + \epsilon \frac{1}{12}\frac{h_1(V_1,V_1)}{(1 + \eta(V_1,V_1))}\bigg)\,.
	\end{align}
	\item The separations built from $V_1$ which make the radar experiment symmetric for the observer $\gamma_1$, c.f. Eq. \eqref{eq:radarsymm},
	\begin{align}
	f^{(0)}(V_1,U_1)=0,\quad f^{(1)}(V_1,U_1)=- \frac{1}{6}h_1(V_1,U_1)+\frac{1}{8}\frac{1}{\sqrt{-\eta(V_1,V_1)}}( \sigma \sqrt{B(V_1,U_1)} - \tilde \sigma \sqrt{\tilde B(V_1,U_1)} )\,,
	\end{align}
	where in the expansion of the $B$ and $\tilde B$ terms $\ell^{(0)}_U(V_1) = \tilde \ell^{(0)}_U(V_1) = \sqrt{-\eta(V_1,V_1)}$.
	For these the time dilation becomes
	\begin{align}\label{eq:time_dilation3} 
		t_2 = t_1 \sqrt{1 + \eta(V_1,V_1)} \bigg(1 + \epsilon \frac{1}{12}\frac{h_1(V_1,V_1)}{(1 + \eta(V_1,V_1))} + \epsilon \frac{ \sigma \sqrt{B(V_1,U_1)} - \tilde \sigma \sqrt{\tilde B(V_1,U_1)}}{8\sqrt{- \eta(V_1,V_1)}(1 + \eta(V_1,V_1))}\bigg)\,.
	\end{align}
\end{itemize}

The zeroth order result yields the special relativistic time dilation with the transformation factor $\sqrt{1 +\eta(V_1,V_1)}$ The correction due to WPE then depends on how the relative velocity between the two observers, whose time dilation we calculate, is determined. In metric electrodynamics the two conditions, having a symmetric radar experiment and a spatial relative velocity coincide, for a general WPE this is not the case. Note that the two different possibilities described in the Eqs.\ \eqref{eq:time_dilation2} and \eqref{eq:time_dilation3} are not different experiments but rather as discussed in Sec.\ \ref{ssec:radarlength} different evaluations of the same experiment, where different $X$ are associated with the measured radar length, cf. Fig.\ \ref{fig:radarsk}. Which one is chosen depends solely on the operational accessibility of the underlying conditions.

Note that there is one further subtlety, when interpreting the Eqs. \eqref{eq:time_dilation2} and \eqref{eq:time_dilation3}. The metric norm of the relative velocity $\eta(V_1, V_1)$ may in general not be an experimentally accessible number, since the radar experiment, with which one may determine the relative velocity, is described by the radar length, which is not given by the Minkowski metric norm. A physical accessible number, we may associate to $V_1$ is one of its radar lengths $\mathcal{L}_{U_1}(V_1)_{\sigma\tilde{\sigma}}$, which we can introduce in the time dilation formulae using Eq. \eqref{eq:radar}:
\begin{align}\label{eq:metrictoradarnorm}
	\eta(V_1,V_1) = - \frac{\mathcal{L}_U(V_1)^2}{4} + \frac{\epsilon}{4}\bigg( \Big( \sigma \sqrt{B(V_1,U_1)}+\sigma \sqrt{\tilde B(V_1,U_1)} \Big) -  2 h_1(V_1,V_1) - h_1(U_1,U_1)\mathcal{L}_U(V_1)^2 \bigg)\,.
\end{align}
Using this, both formulae, Eqs. \eqref{eq:time_dilation2} and \eqref{eq:time_dilation3}, depend on the second order of the Fresnel polynomial as well as on the polarizations of the light used to perform the radar experiment, i.e. $\sigma$ and $\tilde\sigma$. 

Experiments searching for corrections of the time dilation are the Ives-Stillwell and -- in combination with corrections to the length contraction -- the Kennedy-Thorndike type of experiments, which we will describe with the methods developed in this article in an upcoming publication.

\subsection{Length contraction}\label{ssec:lcon}
Consider an object, i.e.\ a separation vector $X_2$, whose endpoints travel on the observer worldlines $\gamma_2$ and $\gamma_3$ with the $L^*$ normalized constant tangents $U_2$ and $U_3$, respectively. This means $X_2$ is at rest and spatial with respect to $U_2$ and $U_3$. We could relax this assumption and calculate the length contraction for different relations between $U_2$ and $X_2$, for example the symmetric radar length relation. However, the calculation of the length contraction of a spatial distance is most compact and most convenient for the comparison of the result with special relativity.

Now consider an observer on a worldline $\gamma_1$ with $L^*$ normalized tangent $U_1$ crossing $\gamma_2$. At this moment, the observer $U_1$ can determine its equal time surface and its spatial directions. They do in general not coincide with the ones determined by $\gamma_2$. The separation vector $X_2$ can then be decomposed into a spatial vector $X_1$ of $U_1$, which represents the separation between $\gamma_1$ and $\gamma_3$:
\begin{align}
X_2 = \beta(X_1,U_2) U_2 + X_1\,.
\end{align}
The aim now is to express the radar length $\mathcal{L}_{U_2}(X_2)$ the observer on $\gamma_2$ associates to $X_2$ in terms of $U_1$ and $X_1$ and compare it to the radar length $\mathcal{L}_{U_1}(X_1)$ the observer on $\gamma_1$ associates to $X_1$.

We start by determining $\beta$ using the spatial condition Eq.\ \eqref{eq:spatial} between $U_2$ and $X_2$, expressed in the general perturbation language of \eqref{eq:Lstarsecorder}, together with the normalization ~$L^*(U_2)=1$, c.f.\ Eq.\ \eqref{eq:Unorm},
\begin{align}
	\beta(X_1,U_2) = - \eta(X_1,U_2) - \epsilon \frac{1}{6} h_1(X_1,U_2)\,.
\end{align}
To eliminate $U_2$, we introduce, as in the previous section, a relation between the observer worldlines $\gamma_1$ and $\gamma_2$ in terms of their relative velocity $t_{2} U_2 = t_1 (V_1 + U_1)$ and eliminate the proper time intervals by using \eqref{eq:time_dilation2} to obtain
\begin{align}
	U_2(V_1,U_1) = \frac{1}{\sqrt{ 1 + \eta(V_1,V_1)}}\bigg(1-\epsilon \frac{1}{12} \frac{h_1(V_1,V_1)}{1+\eta(V_1,V_1)}\bigg)(V_1+U_1)\,,
\end{align}
which implies for $\beta$ 
\begin{align}
	\beta(X_1,U_2(V_1,U_1)) = - \frac{1}{\sqrt{ 1 + \eta(V_1,V_1)}}\bigg(1-\epsilon \frac{1}{12} \frac{h_1(V_1,V_1)}{1+\eta(V_1,V_1)}\bigg) \bigg( \eta(X_1,V_1) + \epsilon \frac{1}{6} h(X_1,X_1)\bigg)\,.
\end{align}
Having obtained these identities, we can express $X_2$ in terms of $X_1,V_1$ and $U_1$ and find the result of the radar experiment in the rest frame of $X_2$ as function of these quantities
\begin{align}
	&\mathcal{L}_{U_2(V_1,U_1)}(X_2(X_1,V_1,U_1))_{\sigma\tilde{\sigma}}\nonumber \\
	&= 2 \sqrt{\frac{\eta(X_1,V_1)^2}{1 + \eta(V_1,V_1)}-\eta(X_1,X_1)} \bigg( 1 + \frac{\epsilon}{2} \frac{C(X_1,V_1,U_1)_{\sigma\tilde{\sigma}}}{12 (1+\eta(V_1,V_1))(\eta(X_1,X_1)(1+\eta(V_1,V_1))-\eta(X_1,V_1)^2)} \bigg)\,, 
\end{align}
with
\begin{align}
	&C(X_1,V_1,U_1)_{\sigma\tilde\sigma} \nonumber\\
	&= 3\bigg(2 h_1(X_1,X_1) - \sigma \sqrt{B(U_2(V_1,U_1),X_2(X_1,V_1,U_1))} - \tilde{\sigma}\sqrt{\tilde B (U_2(V_1,U_1),X_2(X_1,V_1,U_1)) } \bigg) (1 + \eta(V_1,V_1))^2\nonumber\\
	&- 4 \bigg( \eta(X_1,V_1) \Big( 2 h_1(X_1,U_1) + 3 h_1(X_1,V_1) \Big) + 3 \eta(X_1,X_1) \Big( h_1(U_1,U_1) + 2 h_1(V_1,U_1) + h_1(V_1,V_1) \Big) \bigg)(1 + \eta(V_1,V_1))\nonumber\\
	&+ 2 \eta(X_1,V_1)^2 (8 h_1(U_1,U_1) + 16 h_1(V_1,U_1) + 9 h_1(V_1,V_1))\,,
\end{align}
where $B$ and $\tilde B$ must be evaluate at $\ell^{(0)}_{U_2}(X_2) = \tilde \ell^{(0)}_{U_2}(X_2) = \sqrt{-\eta(X_2,X_2)}$, since $X_2$ is spatial with respect to~$U_2$. The above equation is the general length contraction formula implied by WPE. It includes effects from the normalization for the observers with respect to $L^*$ as well as the modified propagation of light determined by $L^\#$ and the possibility of performing the radar experiment with different types of light.

The equation simplifies for the case that the object and the observer are moving in the same direction, i.e. $X_1 = \lambda V_1$. Then $\eta(X_1,X_1) = \lambda^2 \eta(V_1,V_1)$ and $\eta(X_1,V_1) = \lambda \eta(V_1,V_1) = \sqrt{ - \eta(V_1,V_1)} \sqrt{-\eta(X_1,X_1)}$ yields
\begin{align}
	\mathcal{L}_{U_2(V_1,U_1)}(X_2(X_1,V_1,U_1))_{\sigma\tilde{\sigma}} = 2 \sqrt{\frac{-\eta(X_1,X_1)}{1 + \eta(V_1,V_1)}} \bigg( 1 + \frac{\epsilon}{2} \frac{C(X_1,V_1,U_1)_{\sigma\tilde{\sigma}}}{12 (1+\eta(V_1,V_1))\eta(X_1,X_1)} \bigg)\,.
\end{align}
The length contraction factor picks up the following modification from WPE
\begin{align}
\begin{split}
	&\frac{\mathcal{L}_{U_2(V_1,U_1)}(X_2(X_1,V_1,U_1))_{\sigma\tilde{\sigma}}}{\mathcal{L}_{U_1}(X_1)_{\sigma\tilde{\sigma}}} = \frac{1}{\sqrt{1 + \eta(V_1,V_1)}}\\
	& \times \bigg(1 + \frac{\epsilon}{8\eta(V_1,V_1)}\bigg[2 h_1(V_1,V_1)- 4 h_1(U_1,U_1)\eta(V_1,V_1) - \sigma \sqrt{B(V_1,U_1)} - \tilde \sigma \sqrt{\tilde B(V_1,U_1)} +\frac{C(V_1,V_1,U_1)_{\sigma\tilde{\sigma}}}{3(1+\eta(V_1,V_1))} \bigg]\bigg)\,,
\end{split}
\end{align}
where the reference radar length $\mathcal{L}_{U_1}(X_1)_{\sigma\tilde{\sigma}}$ is taken from Eq.\ \eqref{eq:radarspatialsymm}. From this expression, the usual textbook length contraction formula is easily recognized in zeroth order. 

With this, we derived the two classical relativistic effects from WPE, which included effects from a modification of the observer's clock as well as from modified light propagation.

\section{Discussion}\label{sec:disc}
The current and future highly sensitive realizations of the Michelson–Morley, Ives-Stilwell and Kennedy-Thorndike experiments require a clear modern theoretical description of these classical tests of special relativity and local Lorentz invariance. The building blocks of the experiments are the propagation of light and the measurement of radar lengths. Besides modeling the mentioned experiments, the radar length also provides an operational notion of spatial lengths for observers in general.

We demonstrated that based on recent insights of the relation between the geometric optics limit of field theories and the motion of massive and massless point particles \cite{Raetzel:2010je}, the radar length can be derived from a theory of electrodynamics employing the following constructive algorithm:
\begin{enumerate}
	\item Derive the geometric optics limit of the theory of electrodynamics in consideration.
	\item Derive the Lagrange functions $L^\#$ and $L^*$, which define the motion of massless and massive particles on the manifold, from the geometric optic limit of the field theory.
	\item Use $L^*$ to realize the clock postulate, i.e.\ to identify the proper time normalization of observer worldlines~$\gamma$ by choosing a parametrization of $\gamma$ with $L^*(\dot{\gamma})=1$.
	\item Model the radar experiment by demanding that for a direction $X$ the vectors $N^+ = \ell_{\dot{\gamma}}(X)\dot{\gamma} + X$  and $N^- = \tilde \ell_{\dot{\gamma}}(X)\dot{\gamma} + X$ are the tangents of the light rays of the radar signal, i.e.\ are null-vectors of $L^\#$.
	\item A solution of $L^\#(N^\pm)=0$ defines the radar length $\mathcal{L}_U(X) = \ell_{\dot{\gamma}}(X) + \tilde \ell_{\dot{\gamma}}(X)$ an observer on worldline $\gamma$ associates to an object represented by the vector $X$.
\end{enumerate}
This algorithm generalizes the derivation of the radar length in \cite{Pfeifer:2014yua}, where $L^*=L^\#$ and $\ell_{\dot{\gamma}}(X) = \tilde \ell_{\dot{\gamma}}(X)$ was assumed. Both assumptions are in general too strong as we saw during the derivation of $L^*$ and $L^\#$ in Sec.\ \ref{sec:PartProp} and by analyzing under which conditions the radar experiment is symmetric in Sec.\ \ref{ssec:radarsymmetric}.

Together with the definition of an observer's equal time surface and its spatial directions, the radar length and proper time can then be used to derive the classical relativistic observables like time dilation and length contraction, as we did in Sec.\ \ref{sec:classrel}. The only postulate involved in the derivations is the clock postulate, everything else follows from the theory of electrodynamics in consideration. 

To demonstrate this algorithm, we applied it to a general first order modification of metric electrodynamics on Minkowski spacetime we called weakly premetric electrodynamics. The theory we consider includes the electrodynamic sector of the minimal SME, which is usually used as test theory to describe violations of Lorentz invariance, see, e.g.,~\cite{Kostelecky:2002hh}. 

An important feature during derivation we like to highlight is that in order to obtain all contributions to the first order corrections to the light propagation from first order modified electrodynamics, it is necessary to consider the second order geometric optics limit of the theory as displayed in Eq.\ \eqref{eq:FresnelPert}. Only this second order term determines the lightlike directions to first order. For the Lagrangian, which describes the motion of massive particles and observers, the first order in the perturbation suffices. This Lagrangian turned out to be a Finsler function and, thus, we showed that also minimal SME electrodynamics leads to a propagation of particles on Finsler geodesics. This provides one further, so far unobserved, connection between the SME and Finsler geometry.

Our main result is the derivation of the radar length in Sec.\ \ref{ssec:radarlength}. Here, it is most important to notice that two effects of the WPE modify the radar length compared to the result obtained in special and general relativity. The first effect is due to the modified propagation of light, encoded in the Lagrangian $L^\#$, the second, which is often not discussed, comes from the modified proper time normalization of observers encoded in the Lagrangian $L^*$. Both effects contribute to our radar length result~\eqref{eq:radar}. To include both of these effects in the derivation is necessary for the comparison of measurements with the theoretical predictions, since the derivation of the classical relativistic effects inherits this dependence, as we demonstrated in Sec.\ \ref{sec:classrel}.

The whole analysis we performed throughout this article was based on the geometric optic limit of a field theory, here the WPE. These are governed by Hamilton functions, whose level sets represent the dispersion relation of the theory. Thus, our description of observer's measurements yields a consistent description of light propagation as well as observers and their measurements of time and length based on dispersion relations. Hence, in the future we can derive radar length measurements and relativistic observables from modified dispersion relations in general. They are not only investigated in the context of electrodynamics but for example also as effective phenomenology of quantum gravity \cite{AmelinoCamelia:2008qg, Mattingly:2005re,Liberati:2013xla,Barcaroli:2017gvg}.

The precise derivation of observables for the Michelson–Morley experiment based on the formalism developed in this article is work in progress. In the future, the description of the Ives-Stilwell and Kennedy-Thorndike shall follow.
\acknowledgments

The authors thank Manuel Hohmann and Frederic Schuller for insightful and fruitful discussions on relativistic effects. C.P. thanks Steffen Aksteiner for the proper introduction to xAct.
N.G.\ was supported by the German Space Agency DLR with funds provided by the Federal Ministry for Economic Affairs and Energy under grant number 50 OO 1604 and 50 QT 1401. C.P.\  acknowledges support of the European Regional Development Fund through the Center of Excellence TK133 ``The Dark Side of the Universe''.

\bibliographystyle{utphys}
\bibliography{TLV}

\end{document}